\documentclass[sigconf]{acmart}

\renewcommand\footnotetextcopyrightpermission[1]{} 
\makeatletter
\renewcommand\@formatdoi[1]{\ignorespaces}
\makeatother

\usepackage{framed}
\usepackage{balance}

\AtBeginDocument{%
  \providecommand\BibTeX{{%
    \normalfont B\kern-0.5em{\scshape i\kern-0.25em b}\kern-0.8em\TeX}}}

\begin{document}

\title{Data Readiness Report}

\author{Shazia Afzal, Rajmohan C, Manish Kesarwani, Sameep Mehta, Hima Patel}
\affiliation{%
  \institution{\{shaafzal,\hspace{0.2em}rajmohanc1,\hspace{0.2em}manishkesarwani,\hspace{0.2em}sameepmehta,\hspace{0.2em}himapatel\}@in.ibm.com}
}
\affiliation{%
  \institution{IBM Research, India}
}


\begin{abstract}
Data exploration and quality analysis is an important yet tedious  process in the AI pipeline. Current practices of data cleaning and data readiness assessment for machine learning tasks are mostly conducted in an arbitrary manner which limits their reuse and results in loss of productivity. We introduce the concept of a Data Readiness Report as an accompanying documentation to a dataset that allows data consumers to get detailed insights into the quality of input data. Data characteristics and challenges on various quality dimensions are identified and documented keeping in mind the principles of transparency and explainability. The Data Readiness Report also serves as a record of all data assessment operations including applied transformations. This provides a detailed lineage for the purpose of data governance and management. In effect, the report captures and documents the actions taken by various personas in a data readiness and assessment workflow. Overtime this becomes a repository of best practices and can potentially drive a recommendation system for building automated data readiness workflows on the lines of AutoML \cite{feurer2015efficient}. We anticipate that together with the Datasheets \cite{gebru2018datasheets}, Dataset Nutrition Label \cite{holland2018dataset}, FactSheets \cite{arnold2019factsheets} and Model Cards \cite{mitchell2019model}, the Data Readiness Report makes significant progress towards Data and  AI lifecycle documentation.
\end{abstract}

\ccsdesc[500]{General and reference~Evaluation}
\ccsdesc[500]{Software and its engineering~Documentation}
\ccsdesc[300]{Human-centered computing~Walkthrough evaluations}

\keywords{machine learning datasets, data quality, data pre-processing, documentation, data governance }

\maketitle

\pagestyle{empty}

\section{Introduction}

In recent times, there has been an increased emphasis on the importance of standardized documentation practices in the machine learning community. Of particular relevance are the foundational papers Datasheets, ModelCards and FactSheets  (\cite{gebru2018datasheets,mitchell2019model,arnold2019factsheets}) that highlight these standardization efforts for various AI assets in a systematic manner. Their primary objective is to enable transparent communication and transfer of information between creators and consumers of various AI assets. These efforts are also aimed at bench-marking the key characteristics and evaluations of AI assets to empower auditing and enable informed reuse. 


In this paper, we propose the Data Readiness Report as a documentation of operations conducted during the data preparation, cleaning and quality analysis phases in a typical AI life cycle. At a conceptual level it describes the characteristics of a dataset and calibrates it with respect to various quality dimensions relevant to machine learning for different stakeholders. This apprises data practitioners about the challenges in the data upfront and can potentially reduce iterative data exploration. Here, we begin with highlighting the importance of data profiling and quality analysis as a key step before data enters a ML pipeline. We discuss the shortcomings and inefficiencies in current practices related to data pre-processing and quality assessment. Finally we illustrate the concept of Data Readiness Report as a shareable asset that accompanies the data. The report consists of baseline quality indicators as well as operations performed as part of remediation or transformations. 
We foresee the key benefits of the Data Readiness Report to be as follows:
\begin{itemize}
    \item Serve as a shareable data asset giving a complete overview of data readiness for machine learning tasks
    \item Document the data quality issues identified as part of data quality and readiness assessment process
    \item Bring in standardization by expressing properties of data in terminology that is meaningful to AI
    \item Serve as a repository of data operations and sequence in which they are applied
    \item Reduce repetitive data exploration and analysis thereby shortening model development time
    \item Enable auditing of user interactions with data, and therefore, 
    \item Preserve \textit{Who} has modified, transformed, or enriched data in \textit{What} manner at \textit{What} stage
\end{itemize}

We suggest that the Data Readiness Report serves as the missing link between the Datasheets and Model Cards/FactSheets thereby completing the transparency and standardization efforts in the AI documentation pipeline. Throughout the paper the terms data preparation, data cleaning, data quality analysis, etc. are used interchangeably under an umbrella term of data readiness and quality assessment to indicate data operations ordinarily undertaken before data enters a machine learning pipeline.

\section{Motivation}
\begin{figure*}
    \centering
    \includegraphics[width=0.75\linewidth]{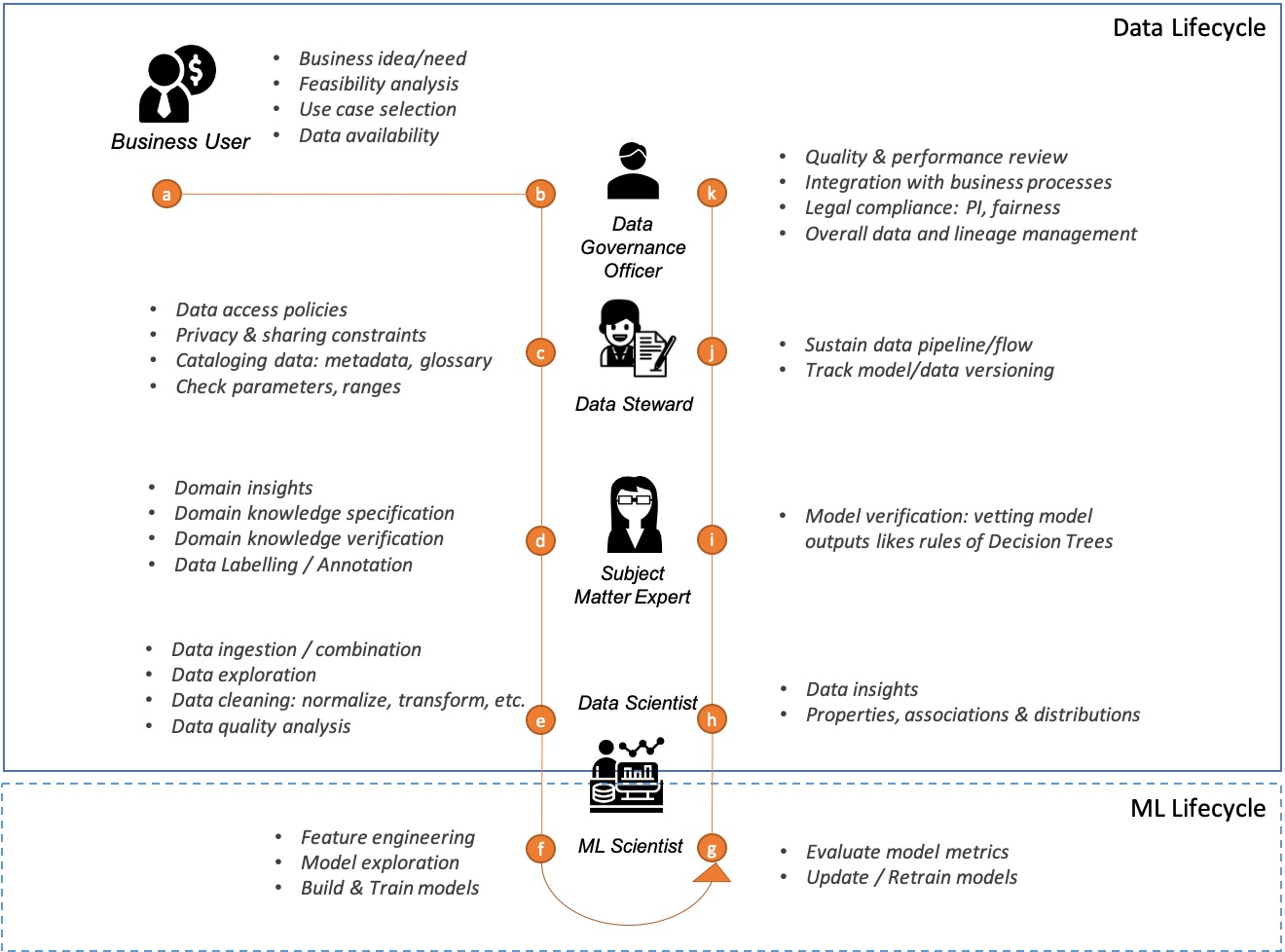}
    \caption{Various personas in an AI pipeline \label{fig:actors}}
\end{figure*}
As machine learning continues to be applied across various domains, the challenges related to data acquisition, its management and control, as well as ensuring its conformance to quality criteria and readiness are also getting multiplied. Data is being ingested in various formats and from diverse sources at remarkable velocity. It is also being consumed and transformed by various technical and non-technical stakeholders at different stages. In a typical AI pipeline, these multiple human actors influence and interact with data in various ways. Figure~\ref{fig:actors} illustrates the role of various personas at different stages of the pipeline. From making the data available for a target use case, to setting access and privacy constraints, checking legal compliance, cataloguing it and getting annotation and labelling along with domain specific insights, are some of the processes that precede the data being acquired by data scientists or machine leaning engineers for use. Post cleaning and pre-processing, models are built and their performance is reviewed. This sets the stage for post production downstream tasks generally centered along review, verification, versioning and integration. The primary objective of this figure is to show that the data ecosystem spans across an organization and human interaction at various levels is an essential component in this workflow. This emphasises the value and need for a transparent and efficient communication of role dependent information between personas. This would not only familiarize them with the profiling or analysis already performed on data but would also enable informed decision-making and infuse efficiency in the process. The Datasheets and Model Cards address such sharing and transparency concerns at data acquisition and model development stages, respectively. Here we focus on the data quality analysis process wherein data is subjected to various cleaning and validation checks before being considered ready for AI tasks.

The journey of data from its collection to serving as training data for an AI system mostly takes place in an adhoc manner and is barely documented.  Anomalies and in-consistencies may creep into the data at any of the ingestion, aggregation or annotation stages. This impacts the quality or readiness of a dataset for building reliable and efficient machine learning models. It is estimated that data scientists spend about $\sim$80\% of their time in iterative pre-processing of data before it is considered to be fit for downstream machine learning tasks. This involves slightly less glamorous operations related to cleansing, validation and transformation of underlying data. Machine learning contexts also introduce additional quality requirements like ensuring adequate representation of target classes, collinearity checks, presence of discriminatory features, occurrence of bias, identification of mislabelled samples and various other data validation checks. These are time consuming and tedious operations but can have a cascading impact on the model design and performance.
\begin{table*}
\caption{A comparison of Data Readiness Report with related work \label{tab:comparison}}
\centering
\begin{tabular}{ |p{2.5cm}|p{1.9cm}|p{1.9cm}|p{1.9cm}|p{1.9cm}|p{1.9cm}|p{1.9cm}| } 
 \hline
 Feature & Datasheets & DataMaid & Codebook & Dataset \newline Nutrition \newline Label & Mithra \newline Label & Data \newline Readiness \newline Report \\ \hline\hline
Basic metadata \newline about dataset & yes & yes & yes & yes & yes & yes
 \\ \hline
Data characteristics & yes & yes & yes & yes & yes & yes
 \\ \hline
Data quality \newline metrics & no & yes & yes & yes & yes & yes
 \\ \hline
Data ethics & yes & no & no & no & yes & yes
 \\ \hline
Task specific & no & no & no & no & yes & no
  \\\hline
Remediations \newline to quality issues & no & no & no & no & no & yes
  \\\hline
Collaborative \newline environment setting & no & no & no & no & no & yes
 \\\hline
Lineage of \newline operations & no & no & no &  no & no & yes
 \\\hline
Nature of output & static \newline questionnaire & overview \newline document & codebook \newline document & interactive \newline report & visual \newline widgets & report \newline document
 \\\hline
Creation method & manual & automated & automated &  automated & automated & automated
\\ \hline

\end{tabular}
\end{table*}
Currently there is no single integrated tooling/solution that systematically addresses these data issues in a streamlined manner. Moreover insights into the characteristics of data and its distribution, that are key to feature engineering, are also conducted in an exploratory and speculative manner. Thus, most effort in data exploration and quality assessment in the data preparation pipeline is dispersed and empirical, thereby not easily transferable across contexts. The knowledge and practice acquires a more experiential and subjective nature which can impact productivity and reuse. Data practitioners may develop affinity to frequently used methods and techniques without conscious regard to maximising value or adopting optimal workflows \cite{rezig2019towards}. This together with a lack of standardized documentation makes the current process inefficient and less productive. The opacity of these operations keep the consumers of the dataset, like machine learning engineers, unaware of the pre-processing or actions already applied on the data. With no supplementary record they have no insight into the quality of incoming data and its readiness for a machine learning task and may spend numerous precious cycles exploring such properties. 

Although Datasheets address the documentation of the data acquisition phase, there is an obvious gap underlining the data preparation and pre-processing phase. We assume a collaborative data quality framework that systematically addresses these issues by providing an integrated platform for data exploration and quality assessment and allows tracking of all user operations and decisions taken by various stakeholders in the pipeline, e.g \cite{rezig2019towards,corrales2018theory}. This not only streamlines the process and optimises productivity but also increases trust and reliability. It can enable creation and cataloguing of operational knowledge in the form of shareable artifacts that describe the data and operations conducted on it. Tasks that require collaboration between personas like a data steward and a subject matter expert can also be conducted in a streamlined manner with specific changes appropriately documented. We propose a Data Readiness Report as an artifact that serves as a certification of the baseline quality of ingested data as well as a record of operations and remediations performed on the data. Unlike Datasheets that focuses on data provenance, the aim here is to account for and record detailed statistical characteristics of data along with the operations conducted on it as part of data quality and readiness for AI modelling. The readiness report becomes a one stop lookup for understanding the quality of data and the transformations it has undergone. The following sections describe the components and use cases of such a report in the context of existing tools and practices.  

\section{Related Work}
Focus on standardized documentation to accompany AI assets is a fairly recent phenomenon. In this section we review prior art related to this theme to show how the proposed readiness report builds on and provides a holistic account of all the operations by data workers in a data quality and readiness pipeline. 

One of the earliest significant works that highlighted this practice was the concept of Datasheets\cite{gebru2018datasheets}. This was proposed as an accompanying record for datasets to document the motivation, composition, collection process, recommended uses, dataset distribution and maintenance etc. thus encouraging transparency and accountability for data used by machine learning community. Data Statements\cite{bender2018data} is similar to Datasheets but concentrated on linguistic datasets focusing specifically on NLP related categories.

Model Cards\cite{mitchell2019model} were developed on the same principle as short documents that could accompany trained machine learning models and provide information about the models like general details, intended use cases, evaluation data and metrics, model performance measures etc. Again the motivation was to increase transparency between various stakeholders of machine learning models and systems.
Similar to model cards but oriented towards AI services are FactSheets\cite{arnold2019factsheets}. A FactSheet for AI Services contains  various attributes of an AI service like intended use, performance, safety, and security etc. It tries to bridge the expertise gap between the producer and consumer of an AI service by communicating the various attributes of the AI services in a standardized way. 

The Dataset Nutrition Label\cite{holland2018dataset} proposes a nutrition label concept for datasets to assess the quality and robustness of a dataset before it is used for AI model development. It provides a comprehensive overview of the core components of a dataset in a standardized format through interrogation of the dataset via various flexible qualitative and quantitative modules. The prototype modules discussed are about metadata, provenance, statistics, variables, pair plots, probabilistic model etc. 

MithraLabel\cite{sun2019mithralabel} generates nutritional labels for datasets in the form of a set of task-specific flexible visual widgets.
It helps determine the fitness of the dataset for specific data science tasks like classification, clustering and ranking. The widgets provide information about representation of minorities, bias and correctness in addition to some general information about the dataset, value distributions, summary sketch, missing values, outliers, etc.

Datamaid\cite{Petersen_Ekstrom} provides automated tools for data quality assessment in R. It auto-generates human-readable assessment overview documents containing data cleaning summaries. It does not provide any tools for fixing issues in the data. Codebook\cite{arslan2019automatically} is another R package that generates rich documentation of the dataset containing metadata, data summary statistics, distribution plots, missing values etc. It generates human and machine-readable metadata which is easy to share.

Lawrence\cite{lawrence2017data} proposes the use of data readiness levels as a common language for assessing the quality and preparedness of a
data set for better project management. It proposes an initial set of descriptors for data readiness in the form of three levels representing different stages of data readiness like accessibility of the dataset, faithfulness and representation of the data, appropriateness of the data set for the context. Castelijns et. al.\cite{castelijns2019abc} present a similar framework focused on machine learning.

The major focus of these works has been on highlighting the data characteristics and data quality issues in specific ways. They do not take into account the remediations to the identified quality issues and explanations for the same. They also do not capture the lineage of data assessment operations and the role of various personas in a collaborative data preparation environment. Table \ref{tab:comparison} encapsulates the positioning of data readiness report with respect to the previously mentioned prior art along various dimensions. The data readiness report not only provides a more comprehensive and holistic view of the data quality issues but it also encompasses other critical information from the data assessment lifecycle like remediations applied, lineage of operations and the actions of various personas during data assessment. The report generation process is largely automated except for the basic metadata module which relies on information outside the context of the data quality pipeline.

\begin{figure*}
    \centering
    \includegraphics[width=0.9\linewidth]{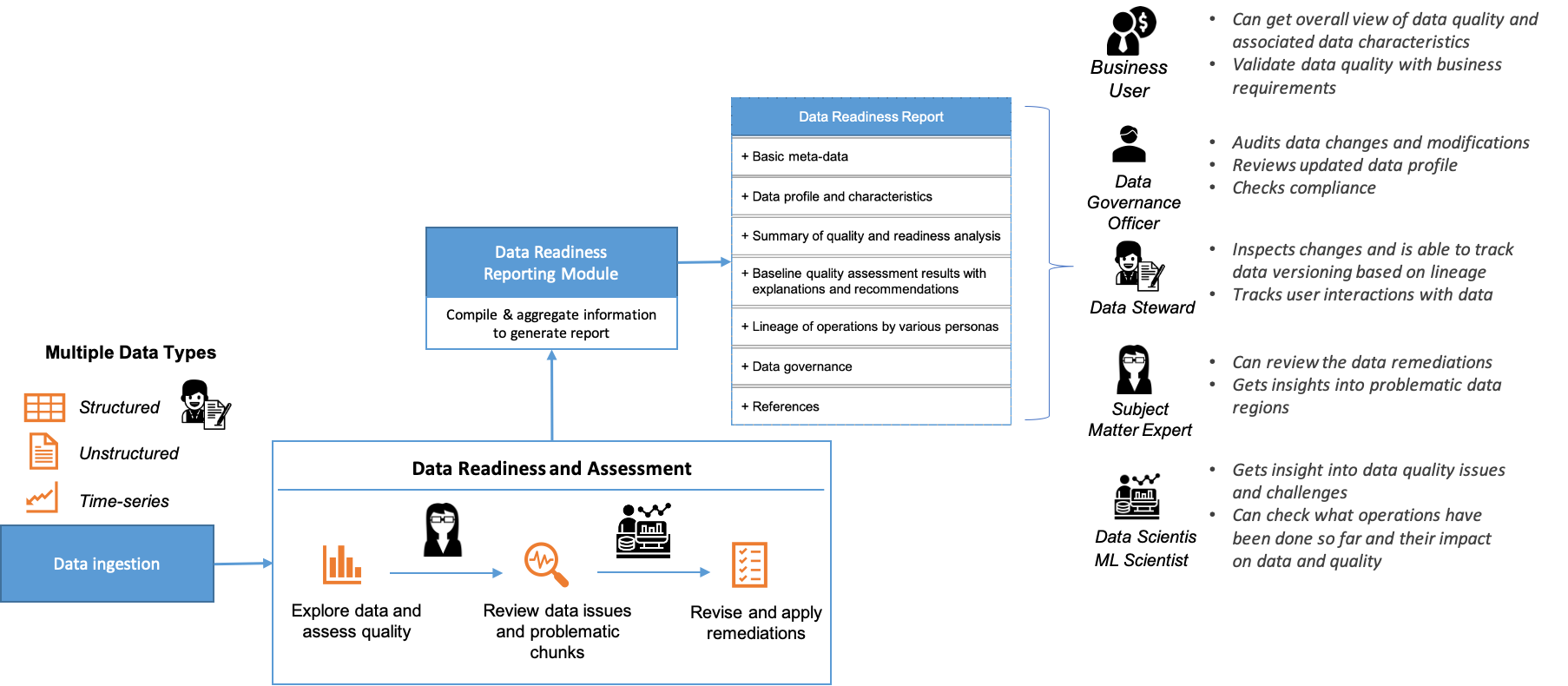}
    \caption{Generation of Data Readiness Report and its Utility Across Personas \label{fig:report_usage}}
\end{figure*}

\section{Data Readiness Report}


Figure \ref{fig:report_usage} depicts how various data workers like data stewards, subject matter experts and data scientists are involved at different stages in the data ingestion and processing workflow. Their interaction with data is determined by their role in the overall pipeline of model development and each may have different function with respect to the kind of input they provide. A data steward may be concerned with bias present in the data, adherence to business constraints, acceptable parameters and ranges for attributes and also monitoring of lineage. An SMEs role could be to validate the annotation quality, identify label noise, review various  recommendations and also offer remedial actions to remove the anomalies. Similarly a data scientist may explore the statistical properties of data and look at other dimensions of quality like outliers, class imbalance, missing values etc. However this involvement of various stakeholders in current practice occurs in an adhoc manner often involving trial and error based iterative workflows. This reduces the scientific rigor of this exercise and also results in loss of productivity as expert human knowledge remains tacit and non-transferable.

In order to increase efficiency and learn best practices over time for driving automation, we introduce the concept of a Data Readiness Report that enables sharing and communication of operations that are done on data to bring it in accordance with the requirements of AI models.  A landscape of \lq data readiness\rq{} is generated by describing features and properties of data along with details of operations applied on it, using terminology that is meaningful to AI. The report serves to quantify the data insights and issues in a format that allows informed use of data. We conceptualise the term \lq data readiness\rq{} to describe its fitness for  machine learning and AI projects.  As such data readiness  is defined in terms of dimensions or attributes that measure or describe the characteristics of data that directly or indirectly impact model development. This includes the data operations undertaken to get data insights, explore features and their associations, measure data quality through validation checks, detect deficiencies and anomalies, and get an estimation of various other criteria that are relevant to data or ML practitioners.  Moreover, the remediations and transformations done to clean and enrich the data and make it ready for AI also contribute to the data readiness framework. 

The Data Readiness Report in itself serves as a documentation of all the quality and readiness analysis including the lineage of transformations applied. It calibrates the quality of a dataset against various dimensions relevant to machine learning. By certifying the dataset in terms of various quality indicators or metrics, it is aimed to be a reference point for a user who wants to understand the data readiness of a given dataset in a transparent manner. The quality analysis should be accompanied with a description of each operation and an explanation of the outputs. The report also records the recommended transformations or suggested remediations for rectifying the issues in data and improving its quality. Since it records the lineage of transformations or operations conducted as part of the remediation, a repository of best practices is learned over time that can be used for improvising recommendations and optimizing data workflows. This can further help in building scalable and principled data assessment solutions by finding the right balance between automation and human input with the latter used strategically for reviewing, verification or approval of automatic decisions. 

Figure \ref{fig:report_usage} also shows some examples in which the report can benefit various personas discussed earlier. For instance, a data scientist A can quickly understand the challenges and quality issues in the data from the report without having to repeat similar analysis conducted by say data scientist B. A subject matter expert can review the remediations done on data and learn how data issues affect quality scores. Similarly, data stewards may be interested in tracking user interactions with data and enabling reliable data versioning based on lineage records. Non technical users like data governance officers or business users may be more interested in overview information like data profile and quality indicators amongst others. The figure shows an illustration of the utility of information provided by the Data Readiness Report for different stakeholders. In practice, users may require and consume information from the report depending on their specific task and requirements.


In the remainder of this section, we describe the composition of the Data Readiness Report in the context of a data quality assessment framework. These components represent information that we believe can be helpful to data practitioners. However this is certainly not exhaustive as various personas in a federated enterprise may require a different view or level of detail. 

\subsection{Basic Metadata}
The introductory section of the report should describe details of the dataset to set the context for the remaining sections. It should also provide accessible links to the original dataset for reference along with standard usage information. It may include the following:
\begin{itemize}
\item basic details of the dataset like name, version, generation date, etc.
\item plain text description of what it contains
\item type of the data: structured, unstructured, time-series, etc
\item intended usage
\item source of the dataset and any specific restrictions on usage
\end{itemize}

Note that this section may have some overlap with a datasheet if available. However, in absence of a detailed data description as required in a datasheet, this first section can serve as an effective summary of key information. Since the information contained in this section is likely to be external to the data pipeline, filling this may require manual effort. The purpose of having this section is to give some background of input data for which a report is generated. This is not anticipated to be an extensive exercise but more of a concise lookup of key facts about the dataset.

\subsection{Summary of Quality and Remediation Assessment}
As the name suggests, the primary goal of this section is to give an overview of the quality and readiness assessment performed on the data. As the number of operations and transformations performed on the data become quite large over time, this summary view provides a quick snapshot of the data profile and quality profile of latest data in comparison to the baseline data. Using a visual info-graphic for illustrating this is a good way to make the review process easier and attractive for consumers of the report.

\subsection{Baseline Data Profile}
This section in the report captures basic characteristics and statistical properties of input data. To simplify the presentation, here we assume that the data is of structured type but in general, the data could be structured, unstructured or time-series and the properties and characteristics would be defined relative to the dataset type. For example, in structured data, a data profile section would include aspects like number of rows and columns, description of each column including datatype, minimum value, maximum value, missing data percentage, number of unique values and other statistical measures. Further for each column, the report can also include details of identified patterns, value distributions and data constraints. The purpose of this section is to provide a detailed profile of the original data by exhibiting all the general traits of the data. It also serves as a benchmark to compare against any updates made on the data during the process of quality assessment and associated transformations.

\begin{figure}
\begin{framed}
\begin{center}
    \textbf{Data Readiness Report}
\end{center}
\begin{itemize}
    \item \textbf{Basic Metadata:} Basic information about data
        \begin{itemize}
            \item Data Owner
            \item Version
            \item Generation Date
            \item Type: structured, unstructured or semi-structured
            \item Description
            \item Tags
            \item Intended Usage
            \item Contact Person
        \end{itemize}
    \item \textbf{Summary of Quality and Readiness Assessment:} Overview of original against final data quality 
        \begin{itemize}
            \item Original and updated data profile 
            \item Original and updated quality profile
        \end{itemize}
    \item \textbf{Baseline Data Profile:} Characteristics of original data
        \begin{itemize}
            \item Number of rows and columns
            \item Basic properties of each column including datatype, description, min value, max value, median, standard deviation, missing data percentage, number of unique values etc 
            \item Column data constraints 
            \item Value distributions
            \item Pairwise column correlations
        \end{itemize}
    \item \textbf{Baseline Quality and Readiness Assessment:} Evaluation of data on various quality dimensions along with explanations and recommended remedial actions. e.g. Label Noise, Class Imbalance, Correlation, Data Homogeneity, Outliers, Missing Values, Inconsistent values, Data Bias, amongst others.
    \item \textbf{Updated Data Profile:} Characteristics of data after applying remediations suggested during the quality assessment and readiness analysis phase.
        
    \item \textbf{Updated Quality and Readiness Assessment:} Quality checks on updated data after application of remediations.
        
    \item \textbf{Lineage of Operations:} Record of data operations by various personas 
        \begin{itemize}
            \item Methods used with input and output parameters
            \item Detailed Results
            \item Explanation of results
            \item Recommendations or suggested actions
            \item SME inputs and remediations applied
            \item Changes to dataset
        \end{itemize}
    \item \textbf{Data Governance: } Set of rules and policies for this data
        \begin{itemize}
            \item Source of Data
            \item Usage Restrictions
            \item Policy Restrictions
            \item License
        \end{itemize}
    \item \textbf{References}
        \begin{itemize}
            \item Details of metrics 
            \item Overview of remediations
            \item Cite implementation details
        \end{itemize}
\end{itemize}
\end{framed}
\caption{Data Readiness Report Template}
\label{tab:report_card}
\end{figure}

\subsection{Baseline Data Quality and Readiness Analysis} 
This is the core section of the report that provides details of the data quality and readiness analysis. This would include the various techniques used to assess the data against dimensions or attributes relevant to a machine learning task like noise, missing values, outliers, high dimensionality, imbalanced class, inconsistency, redundancy, etc. \cite{corrales2018theory}. For example, it may include details of how a dataset fares across certain pre-defined quality metrics known to influence model building efforts \cite{jain2020overview}. The specific format of visualizing this information may vary depending on output requirements and constraints. But key information like the methods used, parameters or configurations applied, details of results including appropriate explanations and suggested actions if any, should be documented. We strongly recommend the use of proper explanations in describing the results of various data operations to enable diverse consumers of the report to understand the issues or challenges in input data. This would improve transparency and ensure that readers of the report make meaningful inferences about the quality checks performed on data. Having appropriate recommendations to improve quality scores wherever possible is strongly advised so that users are not just left with a static analysis but are also made aware of possible ways to fix the identified quality issues. 

\subsection{Updated Data Profile}
The raw data profile may undergo changes once remediations or transformations are done to improve the data quality or remove anomalies.  For example, to rectify class imbalance sampling strategies may be used to make the data more balanced which may result in an increase or decrease in the number of rows depending on the resampling method used \cite{kotsiantis2006handling}. Similarly modifications done to resolve column heterogenity may change the value distributions of target columns \cite{dai2006rapid, gulwani2011automating}. All such changes ought to be documented and this section serves the purpose of providing the relevant updates to the data profile.

\subsection{Updated Quality and Readiness Analysis}
In a typical quality analysis scenario, after checking the data quality on raw data, many transformations or remediations are applied to improve the data readiness. This is typically validated by re-running the quality analysis on the updated dataset and ensuring that modifications done to data have in fact been effective. This should therefore be reflected in appropriate documentation to easily compare the impact and value of performing such operations. 
The expectation is that the quality and readiness metrics on the updated data would be superior compared to the baseline assessments.

\subsection{Lineage of Operations}
In a data readiness assessment pipeline, various personas interact with data in different ways for a variety of reasons. This section will provide a detailed documentation of all the data transformations and operations performed by them in chronological order along with timestamps and other relevant details. This digital trail provides a comprehensive record of how the data has evolved due to interactions with humans in the loop. Having a complete record of all the data quality assessments and applied remediations across roles equips data governance process greatly. With quality control across versions, a data scientist can pick any version of the dataset that best suits them based on the quality scores and the task at hand. It also helps in building a repository of best practices for reuse. For example, comparison of reports can serve as a good review of how the data operations affect data quality and drive automation efforts to recommend optimal sequence of transformations and operations.


\subsection{Data Governance}
Every dataset may be accompanied by a unique set of policy and legal constraints which might restrict its accessibility in different operating environments. This section in the readiness report will highlight any such data usage guidelines. For instance if the data is generated in EU and EEA areas, then GDPR guidelines must be followed. In some cases, data owner might outline custom usage policies which must be adhered to.

\subsection{References}  
This final section catalogs the quality metrics or dimensions used in the report along with appropriate references to give details of methods and algorithms used, parameters and configurations applied. The idea behind this is to improve the scientific rigour so that methods and techniques documented in the report can be effectively re-implemented.

\balance 
\section{Examples}
\begin{figure*}
    \centering
    \includegraphics[width=\linewidth,height=\textheight,keepaspectratio]{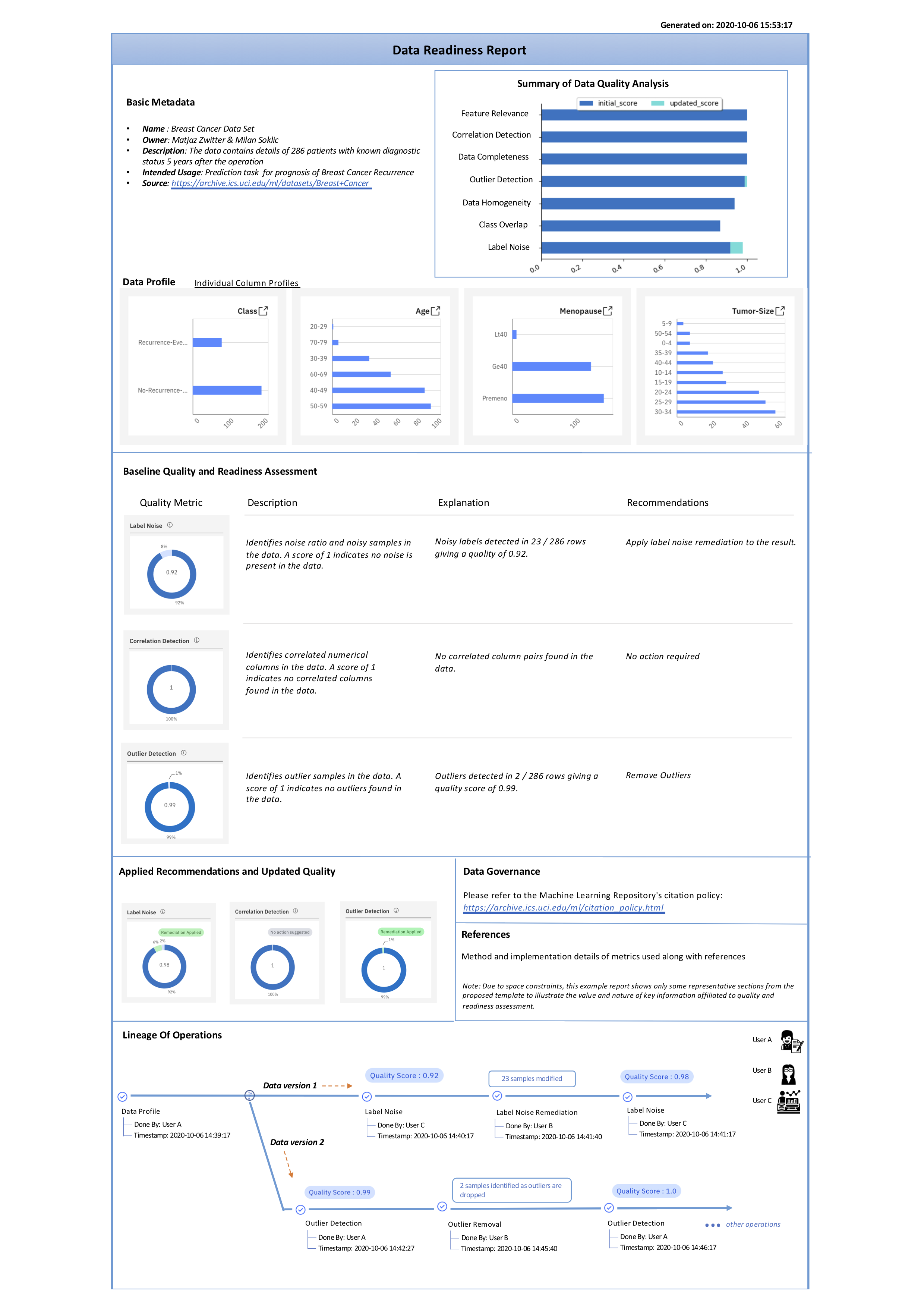}
    \caption{A Sample Data Readiness Report -- Breast Cancer Dataset \label{fig:example_report1}}
\end{figure*}

\begin{figure*}
    \centering
    \includegraphics[width=\linewidth,height=\textheight,keepaspectratio]{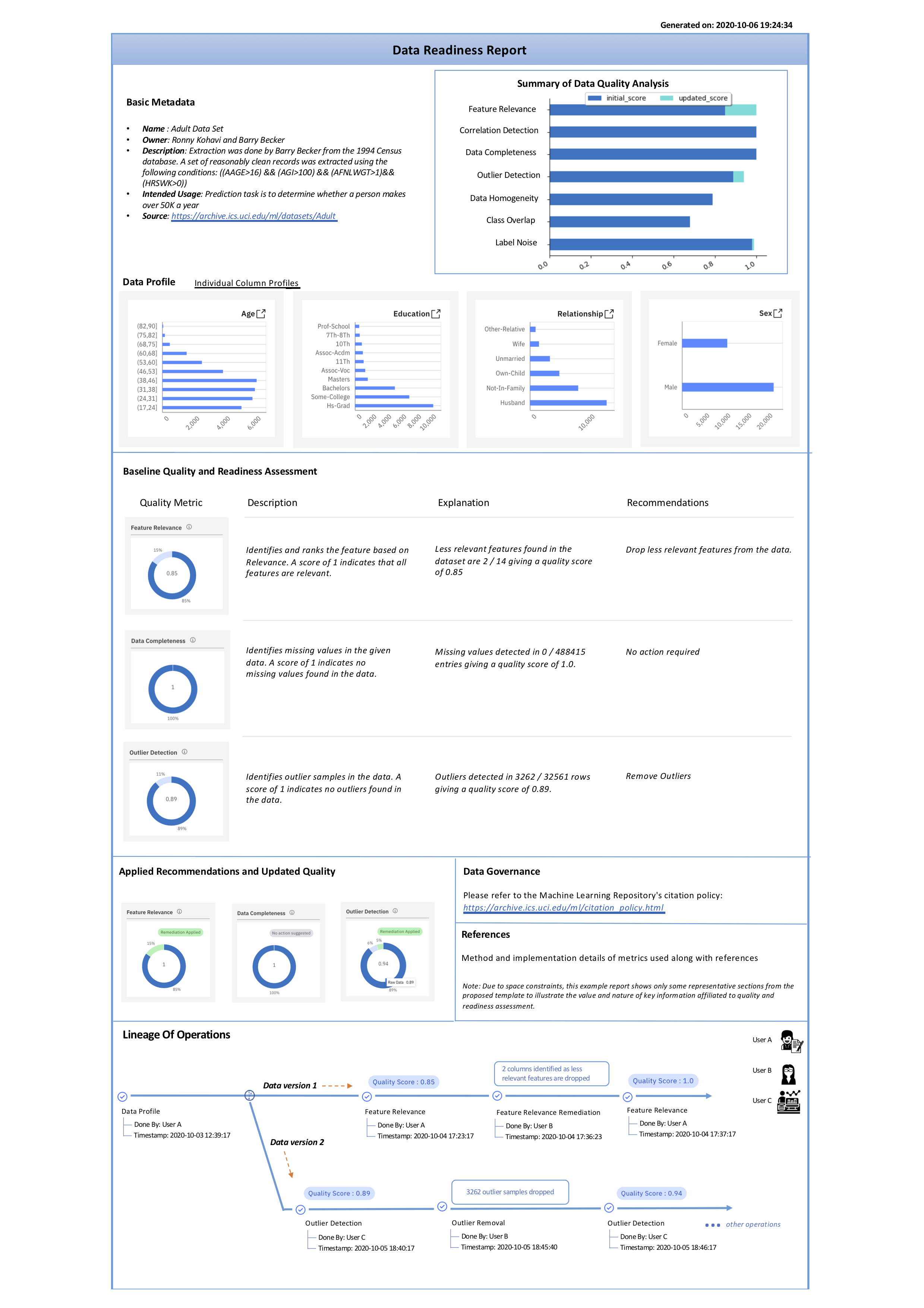}
    \caption{A Sample Data Readiness Report -- Adult Dataset \label{fig:example_report2}}
\end{figure*}

Figures \ref{fig:example_report1} and \ref{fig:example_report2} exhibit sample Data Readiness Reports on the Breast Cancer dataset and Adult dataset respectively from the UCI data repository \cite{Dua:2019}. These mainly contain the key sections covered in the data readiness report template described in Figure \ref{tab:report_card}. We made use of some of the machine learning related quality metrics mentioned in \cite{jain2020overview} for illustration. Due to space constraints, we include only limited features just to exemplify how key information from the quality analysis process can be represented in the report.

\section{Summary}
Data that enters an ML pipeline is subjected to pre-processing by various stakeholders in their own distinctive manner using tools and methods acquired overtime through experience, habit or preference. This adhoc and iterative nature of work limits reuse and results in loss of productivity. Data practitioners also spend a significant percentage of their time in exploring and tackling various data quality issues. This is because they have very limited knowledge about the challenges present in incoming data and whether any modifications or changes have been done to it, and if so, by whom. We introduce the concept of a data readiness report as an accompanying documentation to a data quality and readiness assessment framework that allows data consumers to get detailed data insights into the quality of input data across various standardized dimensions. It serves as a comprehensive documentation of all data properties and quality issues including data operations by various personas to give a detailed record of how data has evolved. Together with the Datasheets and Model Cards this completes the AI pipeline with respect to key factors of transparency, documentation and governance.


\bibliographystyle{ACM-Reference-Format}
\bibliography{sample-base}

\end{document}